\documentstyle[epsfig]{aipproc}
\begin{document}

\title{Testing the Intrinsic Randomnesses in the Angular Distributions
of Gamma-Ray Bursts}
\author{
Attila M\'esz\'aros$^1$, Zsolt Bagoly$^2$,
Istv\'an Horv\'ath$^3$, Lajos G. Bal\'azs$^4$ 
and Roland Vavrek$^4$
}
\address{
$^1$Department of Astronomy, Charles University, V
Hole\v{s}ovi\v{c}k\'ach 2, CZ-180 00 Prague 8, Czech Republic\\
$^2$Laboratory for Information Technology,
E\"{o}tv\"{o}s University, P\'azm\'any P\'eter
s\'et\'any 1/A, H-1518 Budapest, Hungary\\
$^3$Department of Physics, BJKMF, Box 12, H-1456 Budapest, Hungary\\
$^4$Konkoly Observatory, Box 67, H-1505 Budapest, Hungary
}

\maketitle

\begin{abstract}
The counts-in-cells and the two-point angular correlation function
method are used to test the randomnesses in the angular distributions of
both the all gamma-ray bursts collected at BATSE Catalog, and also
their three subclasses ("short", "intermediate", "long"). The methods 
{\it eliminate} the non-zero sky-exposure function of BATSE instrument.
Both tests {\it suggest} intrinsic 
non-randomnesses for the intermediate subclass;
for the remaining three cases only the correlation function method.
The confidence levels are between 95\% and 99.9\%.
Separating the GRBs into two parts ("dim half" and "bright half", respectively)
we obtain the result that the "dim" half shows a non-randomness on the
$99.3\%$ confidence level from the counts-in-cells test.

\end{abstract}

\section*{INTRODUCTION}

Recently, two different articles \cite{hor,muk}
simultaneously suggest that the earlier se\-paration 
of gamma-ray burts (GRBs) - detected by BATSE - 
into short and long subclasses is incomplete. These articles show that
the earlier long subclass alone should be further separated
into a new "intermediate" subclass ($2$ s $< T_{90} < 10$ s) and into
a "truncated long" subclass ($T_{90} > 10$ s). 
Therefore, in what follows, the long subclass will contain only the
GRBs with $T_{90} > 10$ s, and the {\it intermediate subclass} will be
considered as a new subclass ($2$ s $< T_{90} < 10$ s). The "short"
subclass is defined by $T_{90} < 2$ s (for definition of $T_{90}$ see
\cite{mee}).

Fully independently, Bal\'azs et al. \cite{ba98,ba99} 
suggest that GRBs are distributed anisotropically
on the sky. In addition, they show that the short subclass
shows an anisotropy, but the intermediate + long subclasses do not show.
The diffe\-rent behavior is confirmed on 99.7\% confidence level \cite{ba99}.
It is difficult to explain such behavior of subclasses by the instrumental
effects alone. Hence, some {\bf intrinsic} anisotropy should exist. 
A recent study \cite{me}, which is based on the spherical
harmonic analysis, shows that just the GRBs of "intermediate subclass"
{\bf have an intrinsic anisotropy} on confidence level $97\%$.

Here we shortly describe and summarize the new results of two further tests.
The details of these tests will be published elsewhere 
\cite{meb,bag}.

GRBs will be taken between trigger numbers
 0105 and 6963 from Current BATSE Catalog 
\cite{mee} having defined $T_{90}$ (i.e. all GRBs detected up to August 1996
having measured $T_{90}$). From
them we exclude, similarly to \cite{pend,ba98,ba99}, the faintest
GRBs having a peak flux (on 256 ms trigger) smaller than 0.65
photon/(cm$^2$s). 
The 1284 GRBs obtained in this way define the "all" class. From them
there were 339 GRBs with $T_{90} < 2$ s (the "short" subclass), 181 GRBs
with $2$ s $< T_{90} < 10$ s (the "intermediate" subclass) and 764 GRBs with
$T_{90} > 10$ s (the "long" subclass). We will study the all class and
the three subclasses separately. 

\section*{COUNTS-IN-CELLS TEST}

We separate the sky in declination into $m_{dec} > 1$ stripes
having the same "effective" area ($4\pi/m_{dec}$ steradian). This means
that the boundaries
of stripes are the declinations, which ensure that the
probability to have a GRB in a stripe is {\it for any stripe} the same
($=1/m_{dec}$). Because the sky-exposure function of BATSE is {\it not}
depending on right ascension, this may easily be done by the convenient
choices of declinations. 
We also separate the sky in right ascension $\alpha$ into
$m_{ra}>1$ stripes. Hence, 
we separated the sky into $M = m_{dec}\times m_{ra}$
areas ("cells") having the same "effective" size $4\pi/M$ steradian. 

If there are
$N$ GRBs on the sky, then $n =N/M$ is the mean of GRBs 
at a cell. Let $n_i;\; i=1,2,...,M$ be the observed number of GRBs at the
$i$th cell ($\sum_{i=1}^{M} n_i = N$). Then
\begin{equation}
var_M = (M-1)^{-1}\sum_{i=1}^{M} (n_i -n)^2
\end{equation}
defines the observed variance. For the 
given cell structure with $M$ cells, the measured
variance $var_M$
should be identical to the theoretically expected value $n(1-1/M)$. 

$Q$ cell structures may be probed for the same sample of
GRBs. We will choose $m_{dec} = 2, 3, \ldots, 8$
and $m_{ra} = 2, 3, \ldots, 16$. I.e. it will be $Q = 105$.
In the coordinate system with axes
$x = 1/M$ versus $y = \sqrt{var_M/n} = (var/mean)^{1/2}$
the $Q$ values of $(var/mean)^{1/2}$ define $Q$ points, 
where $j=1,2,....,Q$). The theoretical expectation is 
verified by least squares estimation 
(\cite{diggle}, Chapt. 5.3.1.).
Our estimator is the dispersion
\begin{equation}
\sigma_{Q} = \sum_{j=1}^{Q} (y_j - \sqrt{1-1/M}\;)^2\;.
\end{equation}
We throw 1000-times randomly $N$ 
points on the sphere, and repeat the above calculation leading to 
$\sigma_{Q}$ for every simulated sample. Then we compare the size of
the $\sigma_Q$ obtained from this simulation with $\sigma_Q$ obtained
from the actual GRB positions. Let $\omega$ be the number
of simulations, when the obtained $\sigma_{Q}$ is bigger than the
actual value of $\sigma_{Q}$. Then $(100-\omega/10)$ is the confidence level
in percentage.

\section*{TWO-POINT ANGULAR CORRELATION FUNCTION TEST}

Be given $N_D$ GRBs on the sky. There
are $N_D (N_D - 1)/2$ angular distances among them. These measured distances
are binned into bins with binwidth 4 degree.
Then, using a Monte Carlo simulation, there are distributed randomly 
$N_R$ points on the sky, where $N_R
\gg N_D$. (In our case: $N_R = 1000\;N_D$.)
Then the $N_R(N_R-1)/2$ angular distances are binned in the same manner.
In addition, the $N_R N_D$ random-data pairs are also binned in the same
manner. 

In \cite{landy} the following formula is proposed in order to obtain
the $w(\theta)$ correlation function:
\begin{equation}
w(\theta) = \frac{<DD>}{<RR>} - 2  \frac{<RD>}{<RR>} + 1\;,
\end{equation}
where $<...>$ means a normalized mean (see \cite{landy}).

The $1 \sigma$
uncertainty in $w(\theta)$ for the given bin is given by
$\delta w(\theta) = n_{iDD}^{-1/2}$ \cite{landy}.
This formula allows to test the zero expectation
value of $w(\theta)$. For any bin be calculated
the dimensionless value $x$ from the relation
$|w(\theta) - x \delta w(\theta)| =0$. Then $x\sigma$ is the probability
that $w=0$ holds for the given bin.
Because $w=0$ must occur for any bin, simply
the biggest value of $x\sigma$ defines the "Poissonian" confidence level 
\cite{landy}.  

We will verify the confidence levels by Monte Carlo
simulations, too. We will take
- instead of the the actual BATSE positions - randomly generated $N_D$
positions, and then we will repeat the above procedure. 
This simulation will be done 500 times.
If there are $\omega$ such simulations,
when the absolute value of $w_{sim}$ is bigger than the absolute
value of $w_{act}$, then
$(500-\omega)/5$ is the confidence level in percentage.

As the final confidence level the smaller value will be taken.

We exclude the non-uniform sky exposure function
of BATSE instrument as follows.
Assume that the Monte Carlo simulation 
defines a point at a given position. We
generate for any such point an additional random number between
$0$ and $1$, too. 
If this number is bigger than the actual value of sky-exposure
function at that point, then this point is omitted from the Monte Carlo sample.

Note that this test is usually more effective than the counts-in-cells one, 
because usually the "distance-based" tests are more sensitive than "cell-based"
tests (\cite{diggle}, Chapt.2.6).

\section*{THE RESULTS}

The counts-in-cells tests give
$\omega = 287$ ($\omega = 80$, $\omega = 36$, $\omega = 440$)
for all (short, intermediate, long) GRBs. Hence,
{\it the rejection of null-hypothesis of randomness is confirmed 
for the intermediate subclass} 
on the 96.4\% confidence level. For the short and long subclasses, 
respectively, and also for
all GRBs the null-hypothesis cannot be rejected on the
$>95\%$ confidence level.

The calculated four correlation functions give the following results:

1. For the short subclass there is an essential departure from zero
for $\theta = (14 \pm 2)$ degrees. We {\it have} a
99.2\% confidence level for the non-randomness.

2. In the case of intermediate subclass the 
"suspicious" angles are the values
$\theta = (6 \pm 2)$, $\theta = (50 \pm 2)$,
and $\theta = (90 \pm 2)$ degrees. For 
$\theta = (6 \pm 2)$ degrees we {\it have} a
99.8\% confidence level for non-randomness.

3. In the case of long subclass for the angle $\theta = (94 \pm 2)$ degrees 
we {\it have} a 99.0\% confidence level for non-randomness.

4. In the case of all GRBs the following angles are 
"suspicious":  $\theta = (50 \pm 2)$, $\theta = (94 \pm 2)$,
$\theta = (130 \pm 2)$ and  $\theta = (150 \pm 2)$ degrees. For
 $\theta = (94 \pm 2)$ we {\it have} a
99.8\% confidence level for non-randomness.

{\bf FIRST CONCLUSION}: The intrinsic non-random\-ness {\bf is confirmed} on
the confidence level $>95\%$ for {\bf the intermediate subclass} by both
methods; the angular correlation function
(counts-in-cells) method gives a $99.8\%$ ($96.4\%$) confidence
level. This supports the results of \cite{me}.

{\bf SECOND CONCLUSION}: For the remaining two sub-classes and for 
all GRBs 
the null-hypotheses of intrinsic randomnesses are rejected only by the
angular correlation function method on the confidence levels between
$99\%$ and $99.9\%$. The counts-in-cells test, similarly to \cite{me},
did not reject the null-hypothesis of randomness for these cases.

The peak flux = 2 photons/(cm$^2$s) (on $0.256$s trigger) is 
practically identical to the medium. Therefore, we consider
the GRBs having smaller (bigger) peak flux than 2 photons/(cm$^2$s)
as the "dim" ("bright") subclass ("half") of the intermediate subclass. There
are 92 GRBs (89 GRBs) at the "dim half" ("bright half").

The 105 "var/mean" tests for these two parts give
the results that {\it the "dim half" has an intrinsic non-randomness}
on the 99.3\% confidence level;
the "bright half" can well be random.   
The sky distribution of $92$ dim GRBs is shown on Figure 1.

\begin{figure}[b!] 
\centerline{\epsfig{file=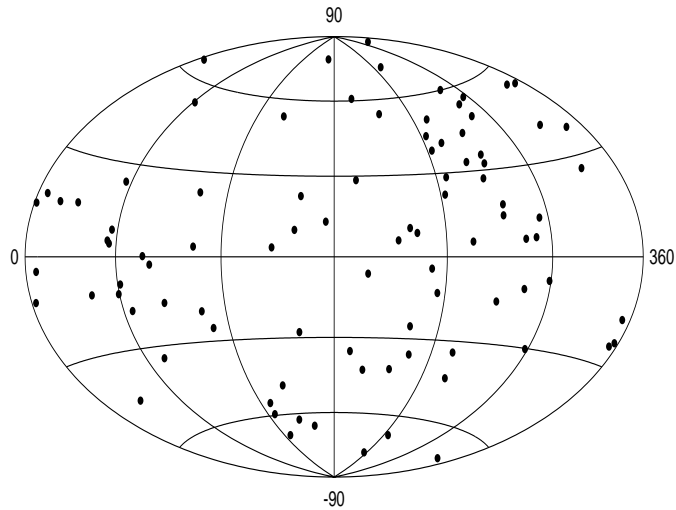,height=3.5in,width=3.5in}}
\vspace{10pt}
\caption{Sky distribution of 82 "dim" GRBs of the intermediate subclass in
equatorial coordinates.}
\label{fig1}
\end{figure}

{\bf THIRD CONCLUSION}: The intrinsic non-ran\-dom\-ness 
of {\bf the "dim half" of the intermediate
subclass of GRBs is confirmed} on the $99.3\%$ confidence level by the 
counts-in-cells method.

\section*{ACKNOWLEDGMENTS}

The authors acknowledge the supports from the following 
grants: GAUK 36/97, GA\v{C}R 202/98/0522, Domus Hungarica
Scientiarium et Artium (A.M.); OTKA T024027 (L.G.B.); F029461 (I.H.).

\end{document}